\documentclass[aps,preprint,amsmath,amssymb]{revtex4}
\usepackage{fmtcount} 
\usepackage{dcolumn}
\usepackage{bm}
\usepackage{epsfig}
\usepackage{amsfonts}

\begin{document}
\title{Generalized Huberman-Rudnick scaling law and robustness of 
$q$-Gaussian probability distributions}

\author{Ozgur Afsar$^{1,}$}
\email{ozgur.afsar@ege.edu.tr}
\author{Ugur Tirnakli$^{1,2,}$}
\email{ugur.tirnakli@ege.edu.tr}
\affiliation{
 $^1$Department of Physics, Faculty of Science, Ege University, 35100 Izmir, Turkey\\
$^2$Division of Statistical Mechanics and Complexity, Institute of Theoretical and 
Applied Physics (ITAP) Kaygiseki Mevkii, 48740 Turunc, Mugla, Turkey
}

\date{\today}

\begin{abstract}
We generalize Huberman-Rudnick universal scaling law for all periodic windows of the 
logistic map and show the robustness of $q$-Gaussian probability distributions 
in the vicinity of chaos threshold. Our scaling relation is universal for the 
self-similar windows of the map which exhibit period-doubling subharmonic 
bifurcations. 
Using this generalized scaling argument, for all periodic windows, as chaos threshold 
is approached, a developing convergence to $q$-Gaussian is numerically obtained both 
in the central regions and tails of the probability distributions of sums of iterates. 
\end{abstract}

\maketitle

\section{Introduction}
\begin{figure}
\includegraphics*[height=8.5cm]{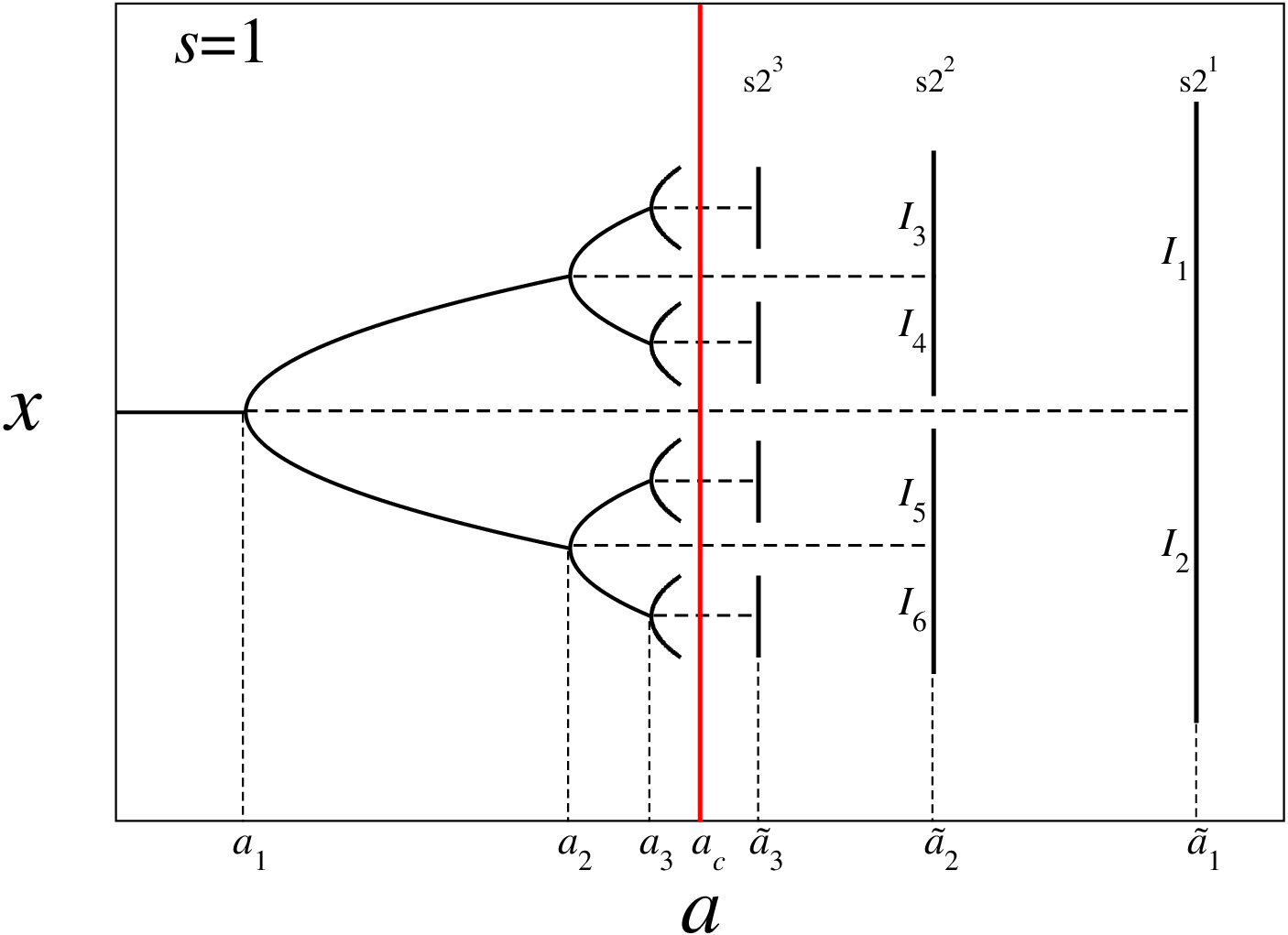}
\includegraphics*[height=8.5cm]{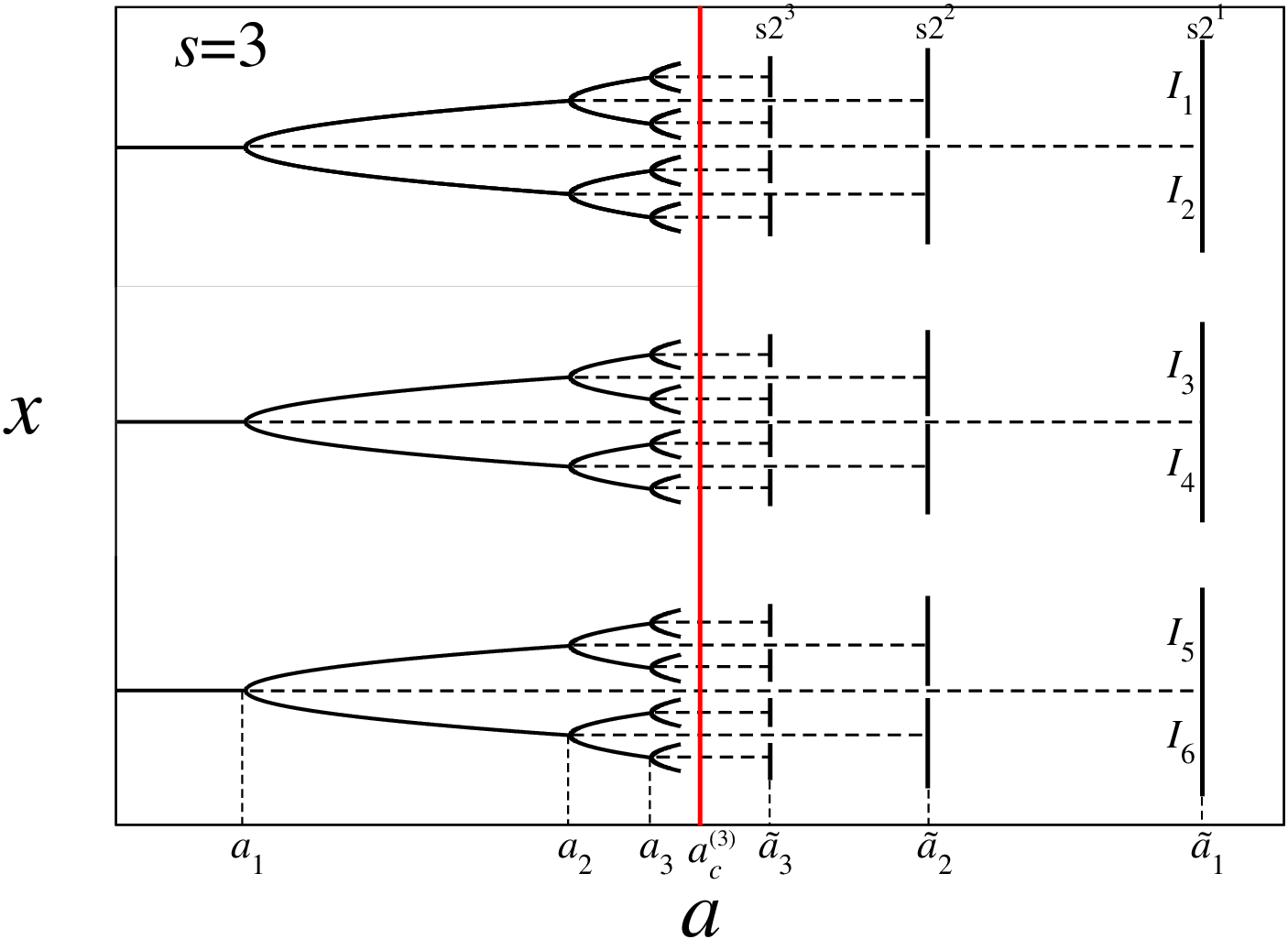}
\caption{
\label{sketchy} 
(a)~Sketchy view of the standard bifurcation diagram of the logistic map. 
Approaching chaos threshold ($a_c=1.401155189...$) from left via period-doubling 
route is evident with bifurcation points denoted as $a_n$ ($n=1,2,...,\infty$). 
Corresponding band merging approach to $a_c$ is depicted with 
merging points denoted as $\tilde{a}_n$ at each of which $2^n$ bands appear. 
(b)~Sketchy view of period~3 window in the chaotic region of the logistic map. 
Approaching chaos threshold ($a_c^{(3)}=1.779818075...$) of this periodic window 
from left via period-doubling route is evident with bifurcation points denoted as 
$a_n$. 
Corresponding band merging approach to $a_c^{(3)}$ is depicted with 
merging points denoted as $\tilde{a}_n$ at each of which $3\times 2^{n}$ 
bands appear.}
\end{figure}

It is well-known that many complex systems exhibit transitions from periodic motion 
to chaos through period doubling route like Rayleigh-Benard system in a 
box \cite{rayleigh}, forced pendulum \cite{humieres}, logistic map \cite{may76}, 
Chirikov map \cite{chirikov79} etc. The logistic map, defined as 

\begin{equation}
x_{t+1}=1-a\, x_t^2  \, ,
\end{equation}
(where $0<a\leq 2$ is the control parameter and the phase space $x_t$ is 
between $[-1,1]$ 
with $t=0,1,2,...$), is a good example to observe Feigenbaum route generated by 
pitchfork bifurcations and its universal features. 
This map has its critical point, denoted by $a_c$, at $a_c=1.401155189...$, which 
can be 
approached from left (i.e., from periodic region) via period doubling procedure where 
$2^{\infty}$ periods accumulate at this critical point usually described as 
chaos threshold. 
This point can also be approached from right (i.e., from chaotic region) via band 
merging procedure where infinite number of bands merge at the critical 
point \cite{crutchfield}. A sketchy view of these approaches from left and from right 
to $a_c$ is given in Fig.~\ref{sketchy}a. 
In the chaotic region ($a>a_c$), there exist many windows of 
higher periodic cascades of $s\,2^n$ periods, where $s$ is an integer and $n$ is the 
degree of period-doublings. It is worth noting that, within each window, 
reverse bifurcations 
of $s\,2^{n+1}$ bands merging into $s\,2^n$ bands can easily be detected although the 
width of these windows decreases rapidly as long as $s\neq1$. 
As an example, $s=3$ case (namely, period~3 case) is illustrated 
in Fig.~\ref{sketchy}b.

This self-similar structure of the map is of sensible importance and leads Feigenbaum 
to develop a scaling theory for the non-chaotic period-doubling region of the 
bifurcations, which enables one to localize the control parameter value at 
the bifurcation from $2^n$ period to $2^{n+1}$ via scaling relation 

\begin{equation}
|a-a_c| \sim \delta^{-n} ,
\label{feigenbaum}
\end{equation}
where $\delta=4.699...$ is the Feigenbaum constant \cite{feigenbaum}. 
On the other hand, it is possible to obtain Eq.~(\ref{feigenbaum}) from the famous 
Huberman-Rudnick scaling law \cite{Huberman}, which shows that the envelope of the 
Lyapunov exponents near $a_c$ exhibits a universal behavior, similar to that of an 
order parameter close to the critical point of phase transition. This relation can 
be written as 
    
\begin{equation}
\lambda = \lambda_0 |a-a_c|^{\nu} ,
\label{huberman}
\end{equation} 
where $a>a_c$, $\nu={\ln2/\ln{\delta}}$, $\lambda$ is the Lyapunov exponent and 
$\lambda_0$ is a constant. 
Eq.~(\ref{feigenbaum}) can easily be obtained from Eq.~(\ref{huberman}). 
For $a$ values slightly above the chaos threshold, there exist $2^n$ 
($n=1,2,....,\infty$) chaotic bands, which approach the Feigenbaum attractor 
as $n \to \infty$ by the band splitting procedure. 
In that region, if we start from two trajectories separated 
by a distance $d_0$ within one of these chaotic bands, 
the separation of trajectories increases exponentially. 
If the trajectories start off in one band, after $2^n$ iterations they will be back 
in the original band. Then they will be separated by the amount

\begin{equation}
d_{2^n}=d_0e^{\lambda2^n}=d_0e^{\lambda_0} ,
\end{equation} 
where $\lambda_0=\lambda2^n$ is the effective Lyapunov exponent 
(a constant value) \cite{Huberman}. Substituting the effective Lyapunov exponent 
into Eq.~(\ref{huberman}) immediately gives 

\begin{equation}
2^{-n} = |a-a_c|^{\ln 2/\ln\delta} ,
\end{equation}
from where Eq.~(\ref{feigenbaum}) is easily obtained.

This scaling relation, in fact, is exactly the one used in \cite{tibet2}, where the 
probability distributions of the sums of the iterates of the logistic map, 
as $a_c$ is approached from the band merging region, have been shown to be 
well approached by $q$-Gaussians provided that the appropriate number of iterations 
($N^*$) is obtained from the above-mentioned scaling relation.

$q$-Gaussians, defined as,

\begin{equation}
P(y) =  \left\{ \begin{array}{ll}
 P(0)\left[1-\beta (1-q) y^2\right]^{\frac{1}{1-q}} &\mbox{for $\beta (1-q) y^2<1$}\\
     0 & \mbox{otherwise}
     \end{array}
     \right.
\label{qGauss}
\end{equation}
(where $q<3$ and $\beta>0$ are parameters and the latter controls the width of 
the distribution) are the distributions that optimize, under appropriate constraints, 
the nonadditive entropy $S_q$ 
(defined to be $S_{q} \equiv \left(1- \sum_i p_i^q\right)/ \left(q-1\right)$), 
on which nonextensive statistical mechanics is based \cite{tccmp,tsallisbook}. 
As $q\rightarrow 1$, $q$-Gaussians recover the Gaussian distribution.

Although in Nature many stochastic processes, consist of sum of many independent
or nearly independent variables, are known to converge Gaussian distribution due 
to the standard central limit theorem \cite{vKa,khinchin}, in recent years several 
complex systems such as low-dimensional dissipative maps in the vicinity of chaos 
threshold \cite{tibet,tsalruiz,afsar-tirnakli}, high dimensional dissipative 
systems \cite{andrea} and conservative maps \cite{queiros1,tassos} are shown 
to exhibit probability distributions that are well approached by $q$-Gaussians.

Our main aim in this paper is two-folded: firstly, we try to generalize the 
Huberman-Rudnick scaling law to all periodic windows of the logistic map, 
secondly, using this generalized version of the Huberman-Rudnick scaling law, 
we analyse the robustness of the probability distribution of the sums of iterates 
of the logistic map as chaos threshold is approached.

\section{Generalization of the Huberman-Rudnick scaling law}

In order to find generalized version of the Huberman-Rudnik scaling law, let us start 
by denoting the accumulation point of a particular periodic cycle $s$ as $a_c^{(s)}$. 
For example, $a_c^{(3)}$ and $a_c^{(5)}$ stand for the accumulation points of period~3 
and period~5 windows inside the chaotic region. These points can be found as 
$a_c^{(3)}=1.779818075...$ and $a_c^{(5)}=1.631019835...\,$. 
Then, one need to check whether the form of the Huberman-Rudnick scaling law is valid 
for all other periodic cycles in the chaotic region. More precisely, one need to check 
whether the exponent $\nu$ in the scaling law is equal to $\ln2/\ln\delta$ as in the 
standard case. Therefore, we have first checked this and found that, for all periodic 
windows, the envelope of the Lyapunov exponents, in $\log\lambda$ vs $\log(a-a_c)$ 
plot, is given by a slope 0.449, which is nothing but $\ln2/\ln\delta$. 
Hence we can now write the Huberman-Rudnick scaling law for all periodic cycles as 

\begin{equation}
\lambda = \lambda_0 \left[a-a_c^{(s)}\right]^{\ln2/\ln\delta} .
\end{equation}
At this point, it should be recalled that, for a particular period $s$ window, 
a trajectory that starts in one band will be back in the same band after $s\,2^{n}$ 
($n=1,2,...,\infty$) iterations. If we use this feature in the definition of the 
effective Lyapunov exponent, namely $\lambda_0=\lambda s 2^{n}$, then one can 
write the Huberman-Rudnick scaling law for other periodic windows as

\begin{equation}
\lambda=\lambda\,s\, 2^{n} \left[a-a_c^{(s)}\right]^{\ln 2/\ln\delta}
\end{equation}
and

\begin{equation}
2^{-n} = s \left|a-a_c^{(s)}\right|^{\ln 2/\ln\delta} .
\end{equation}
This equation enables us to obtain the generalized Huberman-Rudnick scaling law as 

\begin{equation}
\left|a-a_c^{(s)}\right|=\delta^{-n-\frac{\ln s}{\ln2}} .
\label{scaling}
\end{equation}
This new scaling relation is valid for all periodic windows including the standard 
case for $s=1$, which immediately recovers the standard scaling law given 
in Eq.~(\ref{feigenbaum}).

\section{Robustness of Probability Distributions}

Now let us concentrate on the probability distributions of the sums of iterates of the 
logistic map, which can be written as  

\begin{equation}
y:= \sum_{i=1}^N \left(x_i -\langle x \rangle\right) \;\; ,
\label{y}
\end{equation}
where $x_i$ are the iterates of the logistic map and $x_1$ is the initial value 
regarded as a random variable. It has analytically been proved that, for strongly 
chaotic systems, the probability distribution of $y$ becomes Gaussian for 
$N \to \infty$ \cite{billingsley,beck90}. Here, the average $\langle ... \rangle$ is 
calculated as time average. 
On the other hand, as mentioned before in Section I, several complex systems of the 
type low and high dimensional dissipative and conservative exist where the probability 
distribution does not approach to Gaussian, and therefore violating the standard 
central limit theorem due to possible lack of ergodicity and mixing properties. 
For such systems, it is necessary to take the average over not only a large number of 
$N$ iterations but also a large number of $M$ randomly chosen initial values, namely,

\begin{equation}
\langle x \rangle = \frac{1}{M} \frac{1}{N} \sum_{j=1}^{M} \sum_{i=1}^N x_i^{(j)}\; .
\end{equation}  

As chaos threshold is approached, the logistic map has already been studied in this 
respect \cite{tibet,tibet2,robledo1,grassberger}. As it has already been argued 
in \cite{tibet2}, in principle, in order to attain chaos threshold point exactly 
(i.e., approaching this point with infinite precision), one needs to take 
$n\rightarrow\infty$, which, in other words, means that the necessary number of 
iterations to achieve the limit distribution at the chaos threshold is 
$N^*\rightarrow\infty$ since $N^*=2^{2n}$. Since this is, no doubt, unattainable 
in any numerical experiment, one can only approach to this critical point using 
the appropriate values for ($a,N^*$) pairs coming from the Huberman-Rudnick 
scaling law. 
As long as this scaling law is obeyed, developing $q$-Gaussian shape of the 
limit distribution, as chaos threshold is approached, has been clearly shown 
in \cite{tibet2}. 
On the way of approaching chaos threshold, for any approximation level of 
finite ($a,N^*$) pairs, if the number of iterations used is too much larger than 
$N^*$ and therefore violating the Huberman-Rudnick scaling law (i.e., $N>>N^*$), 
it is of course not surprising that the probability distribution starts to 
approach to Gaussian form from its central part since the system starts to feel that 
it is not exactly at the chaos threshold. Such numerical examples can be found 
in \cite{tibet2,robledo1}. 
On the other hand, if the number of iterations used is too much smaller than 
$N^*$ and therefore violating again the Huberman-Rudnick scaling law (i.e., $N<<N^*$), 
then the summation starts to be inadequate to approach the shape of the limit 
probability distribution and it exhibits a kind of peaked or multifractal 
distribution. Such numerical examples have already been given 
in \cite{tibet,robledo1,grassberger}.

In the remainder of this work, we try to provide further evidence on the 
robustness of the $q$-Gaussian probability distributions seen as the chaos 
threshold is approached. In order to accomplish this task, we investigate 
other periodic windows (numerically chosen examples are period~3 and 5) 
making use of our generalized Huberman-Rudnick law.

As separated band structure for periodic cycle $2$ goes from 
$2^1$ to $2^\infty$ with $2^n$ ($n=1,2,...,\infty$), for any periodic cycle $s$, 
the same behavior would be to go from $s\,2^0$ to $s\,2^\infty$ with 
$s\,2^k$ ($k=0,1,...,\infty$).
It is evident that there is $k \to (n-1)$ transformations between $k$ and $n$
mathematically.
Generically, for any periodic cycle $s$ in the chaotic region, one needs to 
perform $s\,2^k$ iterations of the map for a given initial 
value with a control parameter $a$ obtained from the generalized scaling law. 
After $s\,2^k$ iterations, the system will basically fall into 
the same band of the band splitting structure. This means that the sum of the 
iterates $\sum_{i=1}^{s\,2^k} x_i$ will essentially approach to a fixed value 
$w =s\,2^k \langle x \rangle$ plus a small correction $\Delta w_1$ which describes the 
small fluctuations of the position of the $s\,2^k$th iterate within the chaotic band. 
Hence, one can write

\begin{equation}
y_1=\sum_{i=1}^{s 2^k} (x_i-\langle x \rangle ) =\Delta w_1.
\end{equation}
If we continue to iterate for another $s\,2^k$ times, we obtain

\begin{equation}
y_2=\sum_{i=s 2^{k}+1}^{2\, s 2^k} (x_i-\langle x \rangle )= \Delta w_2.
\end{equation}
The new fluctuation $\Delta w_2$ is not expected to be
independent from the old one $\Delta w_1$, since correlations of
iterates decay very slowly if we are close to the critical point.
Continuing this $2^k$ times, we finally obtain

\begin{equation}
y_{2^k}= \sum_{i=s 2^{2k}- s 2^{k}+1}^{s 2^{2k}} 
(x_i- \langle x \rangle )=\Delta w_{2^k}
\end{equation}
if we iterate the map $s\,2^{2k}$ times in total.  
The total sum of iterates

\begin{equation}
y=\sum_{i=1}^{s 2^{2k}} 
(x_i -\langle x \rangle )=\sum_{j=1}^{2^k} \Delta w_j
\end{equation}
can thus be regarded as a sum of $2^{k}$ random variables $\Delta w_j$, 
each being influenced by the structure of the $s\,2^{k}$  chaotic bands at distance 
$a-a_c^{(s)}= \delta^{-n-\frac{\ln s}{\ln2}}$ from the Feigenbaum attractor.
At this distance to chaos threshold, in order to see the limit distribution, 
appropriate number of iterations would be $N^*=s\,2^{2k}$, which 
corresponds to $N^*=s\,2^{2n-2}$  after $k \to (n-1)$ transformation.

\begin{figure}
\includegraphics*[height=9cm]{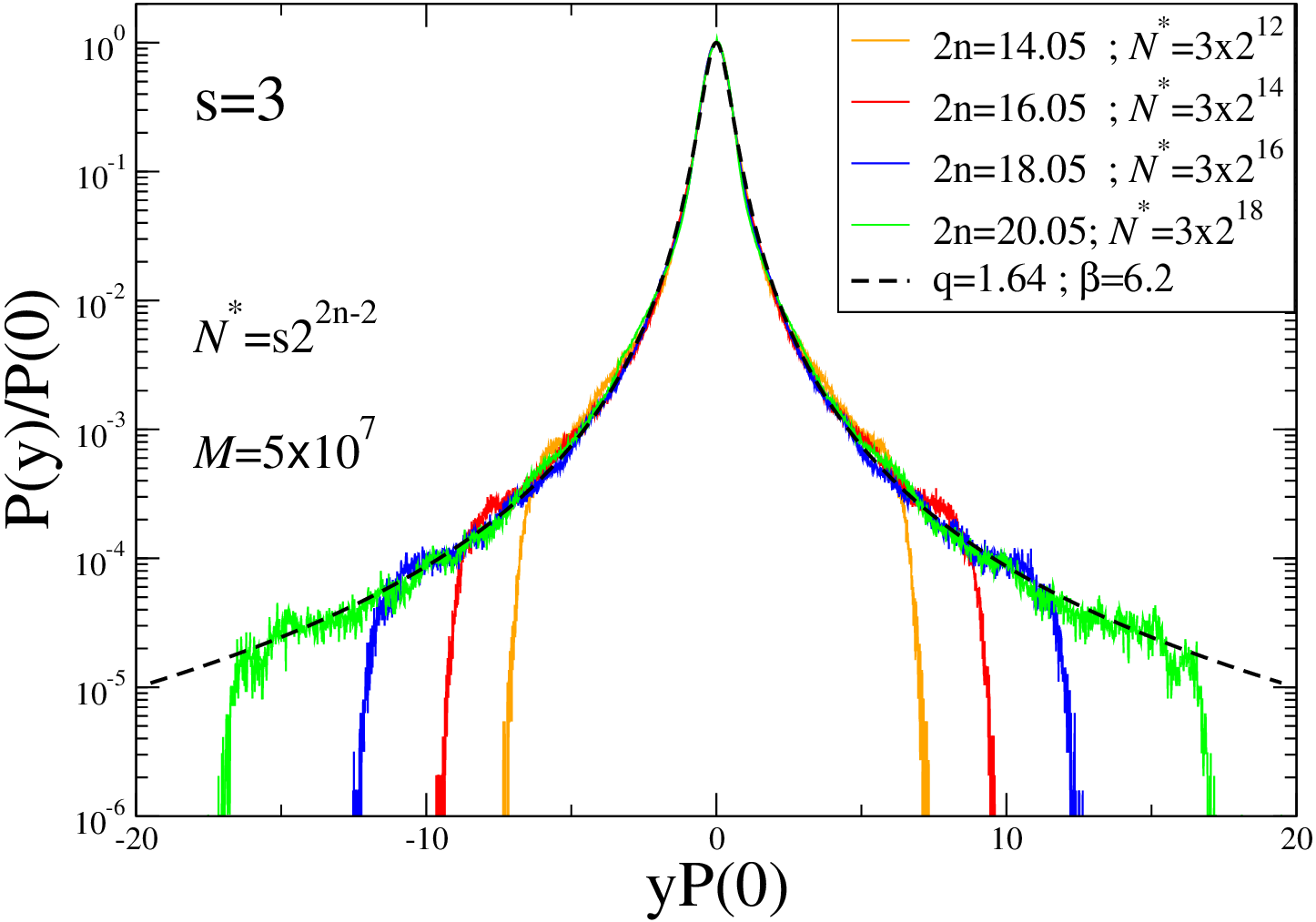}
\includegraphics*[height=9cm]{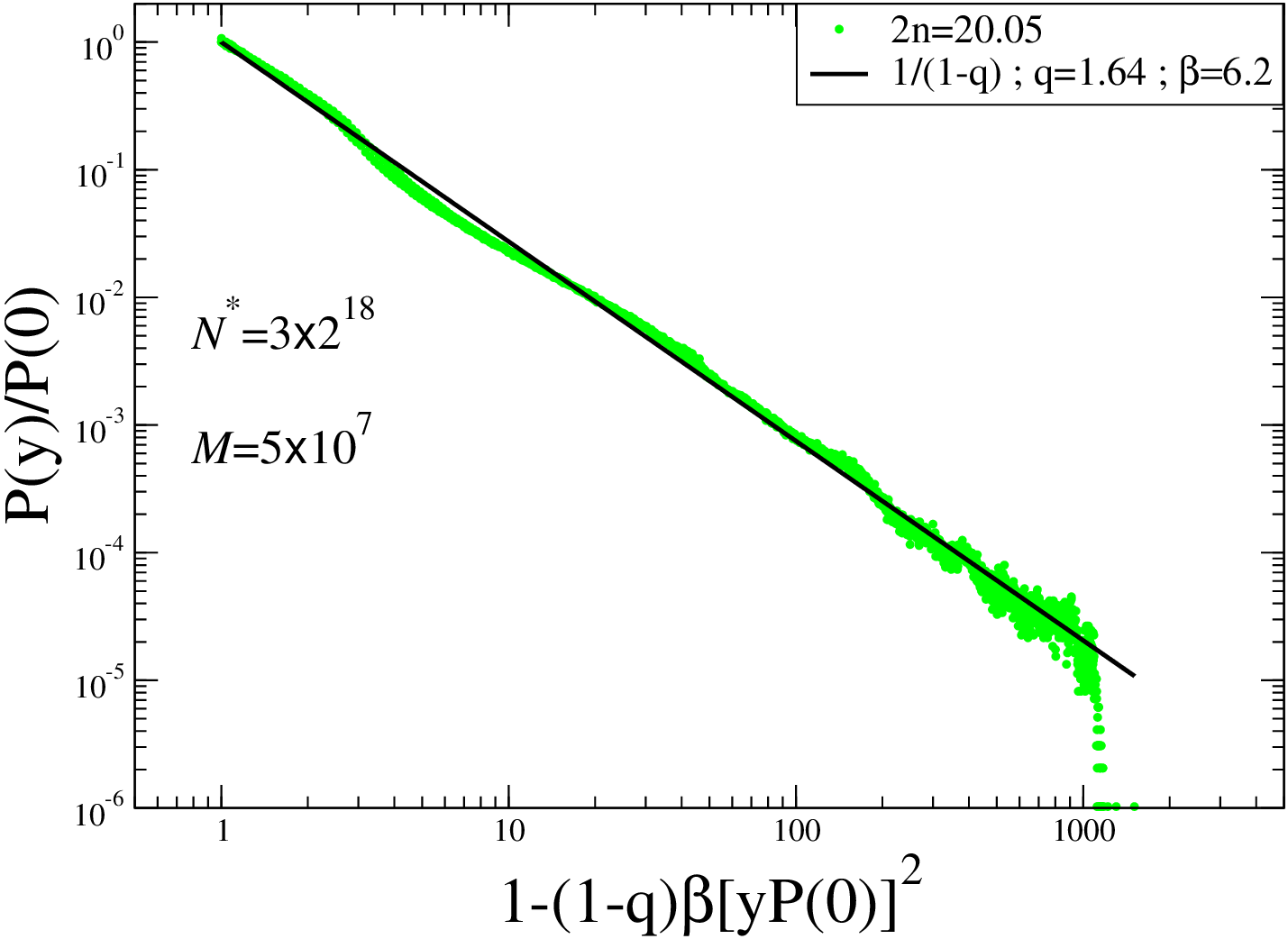}
\caption{\label{fig:Fig 2} (a)~Probability distribution functions of period~3 ($s=3$) 
for four representative cases with different $n$ values. 
As $n$ increases, numerical convergence to a $q$-Gaussian with $q=1.63$ 
and $\beta=6.3$ is appreciated.
(b)~The same distribution plotted against $1-(1-q)\beta[yP(0)]^2$ on a log-log plot 
for the case which is the closest to the chaos threshold.
A straight line is expected with a slope $1/(1-q)$ if the curve is a $q$-Gaussian.
It is clearly seen how the straight line is surrounded by the log-periodically 
modulated curves.}
\end{figure}

\begin{figure}
\includegraphics*[height=9cm]{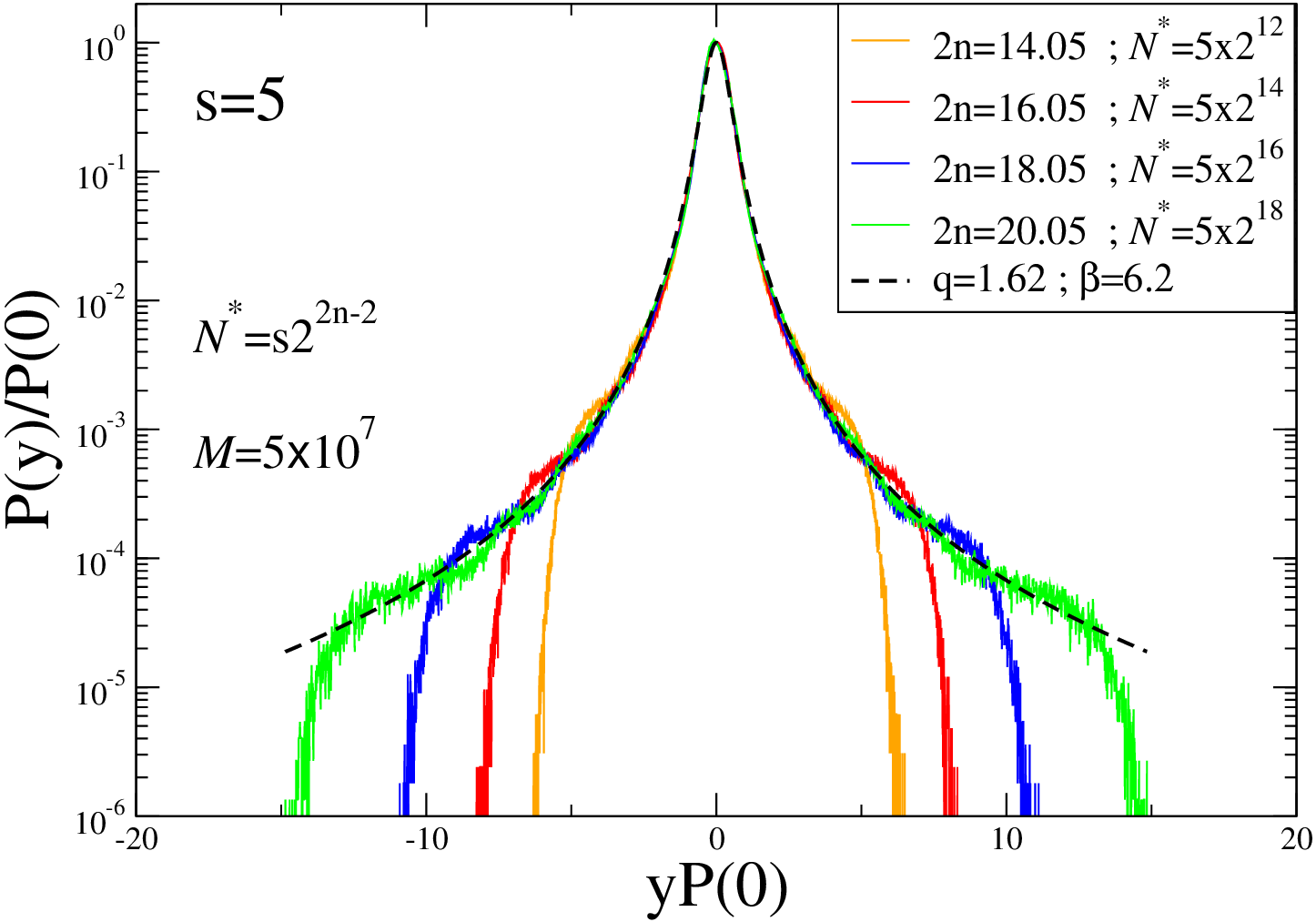}
\includegraphics*[height=9cm]{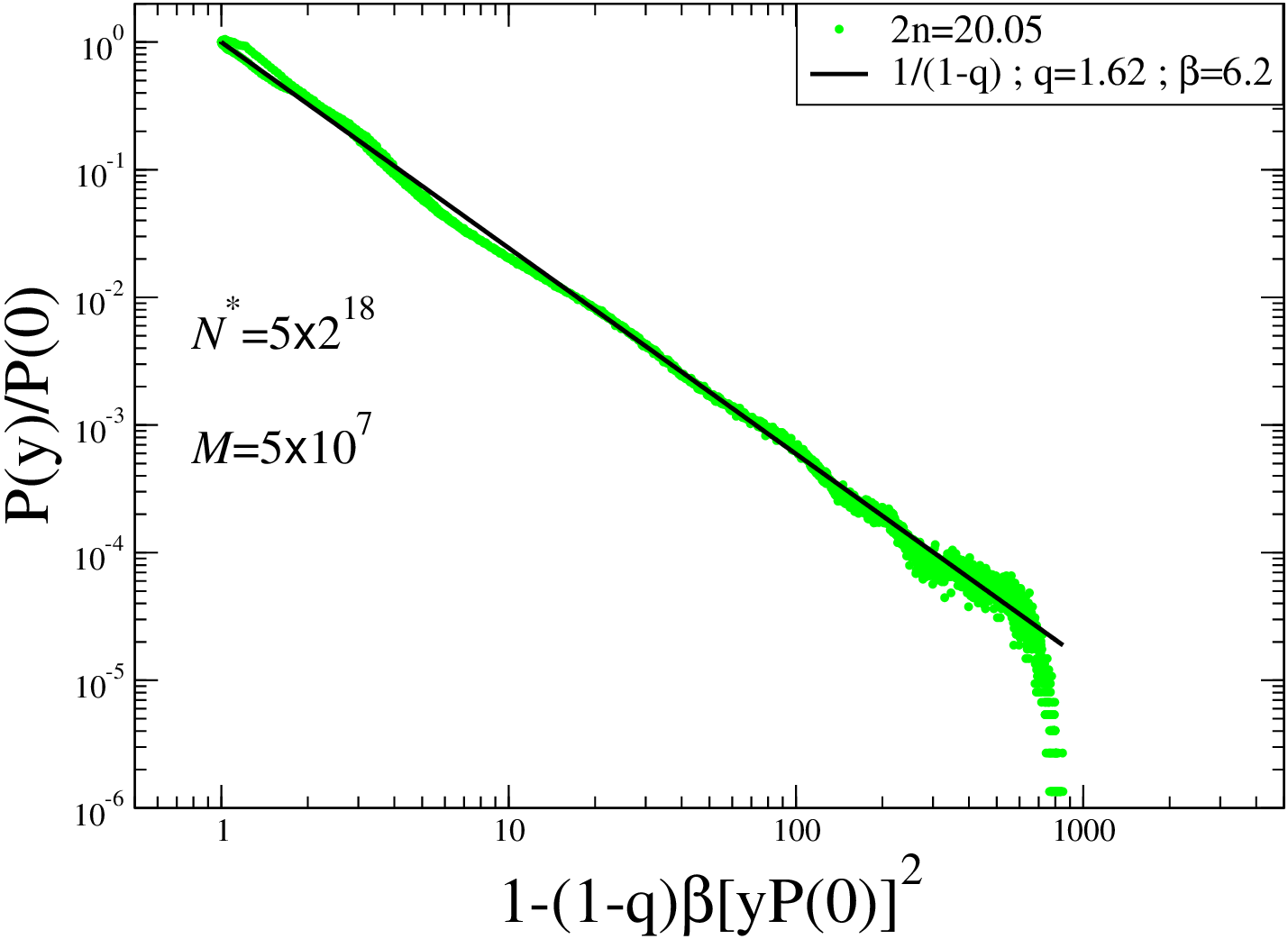}
\caption{\label{fig:Fig 3} (a)~Probability distribution functions of period~5 ($s=5$) 
for four representative cases with different $n$ values. 
As $n$ increases, numerical convergence to a $q$-Gaussian with $q=1.62$ and 
$\beta=6.3$ is appreciated.
(b)~The same distribution plotted against $1-(1-q)\beta[yP(0)]^2$ on a log-log plot 
for the case which is the closest to the chaos threshold. 
A straight line is expected with a slope $1/(1-q)$ if the curve is a $q$-Gaussian.
It is clearly seen how the straight line is surrounded by the log-periodically 
modulated curves.}
\end{figure}

Now we are ready to check the shape of the probability distribution of any periodic 
windows obeying our generalized Huberman-Rudnick scaling law. 
Chosen examples of possible periodic windows are period~3 and period~5 since 
they are the largest two periodic windows available in the chaotic region. 
Although conceptually 
nothing is changed for small sized windows, numerical analysis is getting more 
difficult as windows sizes are decreasing. 
Numerically used values are given in the Table for our period~3 and 5 analysis. 
Control parameter values $a$ are chosen so that the precision of corresponding 
$n$ values, coming from the generalized Huberman-Rudnick scaling law as 

\begin{equation}
n=-\frac{\ln\left|a-a_c^{(s)}\right|}{\ln\delta}-\frac{\ln s}{\ln2} \; ,
\label{nscaling}
\end{equation}
would be the same (see Table). This means that we are approaching the critical point 
with $a$ values located on a straight line with a given slope. 
Our results are given in Fig.~2 for period~3 and in Fig.~3 for period~5. 
In both cases four representative points systematically approaching the 
chaos threshold is given. 
It is clear from Fig.~2a and Fig.~3a that the probability distributions of both 
periodic windows approach to a $q$-Gaussian. It is also evident that, 
as the chaos threshold is better approached, the tails of the distribution 
develops better on the $q$-Gaussian, signaling that the limit distribution obtained 
at the exact chaos threshold point would be a $q$-Gaussian with infinitely long tails. 
We also present the same data in Fig.~2b and Fig.~3b in a different way so that 
a straight line would be expected for $q$-Gaussians. Only the closest cases to the 
chaos threshold for each periodic window are plotted. It is seen from these plots 
that the curves develop on top of a straight line surrounded by log-periodic 
modulations.

\addtocounter{footnote}{1}

\begin{table}[h]
\begin{center}
\caption{Parameter values used in this work. The values of $n$ obtained from 
the generalized scaling law using Eq.~(\ref{nscaling}), the corresponding $N^*$ values 
and the values of $q$ and $\beta$ (estimated from simulations) are listed. 
$s=1$ case, already discussed in \cite{tibet2}, has also been included in the 
Table for comparison. } 
\vspace{0.6cm}
\begin{tabular}{|l|l|l|c|l|l|l|}
\hline \hline
 $s$  &   $a$   &  $2n$   & $N^*$ & $q$   & $\beta$ \\
 \hline \hline
    &  $1.401588$&  10.05    & $2^{10}$ &          &        \\ \cline{2-3}
\cline{2-4}        
  1$^\dagger$ \footnote[0]{$^\dagger$ Values related to $s=1$ case are taken directly 
 from the Table given in ref.~\cite{tibet2}.}  & $1.401248$ &  12.05    & $2^{12}$ & 1.70    &  6.2   \\ \cline{2-3}
\cline{2-4}
     & $1.401175$ &  14.05    & $2^{14}$ &          &         \\ \cline{2-3}
\cline{2-4}
     & $1.40115945$  &  16.05    & $2^{16}$ &          &         \\ \cline{2-3}     
\hline \hline
    &  $1.779819805038384$ &  14.05    & $3\times2^{12}$ &          &        \\ \cline{2-3}
\cline{2-4}         
  3  & $1.779818446177396$ &  16.05    & $3\times2^{14}$ & 1.64     &  6.2    \\ \cline{2-3}
\cline{2-4}
     & $1.779818155150985$ &  18.05    & $3\times2^{16}$ &          &         \\ \cline{2-3}
\cline{2-4} 
     & $1.779818092822039$  &  20.05    & $3\times2^{18}$ &          &         \\ \cline{2-3}     
\hline \hline
    &  $1.63102039110464$ &  14.05    & $5\times2^{12}$ &          &        \\ \cline{2-3}
\cline{2-4}
  5   & $1.63101995463619$ &  16.05    & $5\times2^{14}$ & 1.62     &  6.2    \\ \cline{2-3}
\cline{2-4}
     & $1.63101986115802$ &  18.05    & $5\times2^{16}$ &          &         \\ \cline{2-3}
\cline{2-4}
     & $1.63101984113785$  &  20.05    & $5\times2^{18}$ &          &         \\ \cline{2-3}          
\cline{2-5}
  
\hline
\hline
\end{tabular}
\end{center}
\label{avalues}
\end{table}

\section{Conclusions}

Our main results obtained in this paper can be summarized as follows:
(i)~For the logistic map having self-similar structure, Huberman-Rudnick
universal scaling law has been generalized, which becomes now consistent to all 
periodic windows in the chaotic region of the map. 
This new generalized scaling law is of sensible importance since it enables us to 
produce self-similar structure of the map and to explain all band merging structures 
in all available periodic windows using only one generalized formula. 
(ii)~The standard Huberman-Rudnick scaling law has already been used 
in \cite{tibet2,tsatir} and $q$-Gaussian probability distributions have been observed 
as the standard period~2 accumulation point is approached. 
However, in order to test the robustness of $q$-Gaussian distributions, 
a first straightforward attempt should be to analyse other critical points 
(chaos thresholds) of different periodic windows located in the chaotic region 
of the logistic map. 
Since the generalized Huberman-Rudnick scaling law obtained in the first part of 
this paper now enables us to localize appropriate ($n,N^*$) pairs as the 
accumulation point is approached, we managed to check two representative 
periodic windows. 
For each case studied here (and possibly for all other periodic windows) it is 
numerically shown that the $q$-Gaussian probability distributions with 
log-periodic oscillations are again the observed distributions and developing 
better as the critical point becomes closer. These results clearly indicate 
the robustness of the $q$-Gaussian probability distributions seen in the vicinity 
of chaos threshold. 
Although the obtained $q$ values seem to exhibit a slow decreasing tendency as the 
size of the periodic window becomes smaller, we believe that the genuine limit 
distribution of the chaos threshold (for all $s$ values) would converge to 
a $q$-Gaussian with a unique $q$ value, which is expected to be in the 
interval $[1.6,1.75]$. 

Finally it is worth mentioning that the results obtained here are expected to be 
valid for all other dissipative maps sharing the same universality class with 
the logistic map. 
As an open question that can be addressed in a future work, one can mention 
the analysis of appropriate scaling law for the systems exhibiting 
quasi-periodic route to chaos.

\section*{Acknowlegments}
This work has been supported by TUBITAK (Turkish Agency) under the Research 
Project number 112T083.


\end{document}